\documentclass[a4paper,12pt]{article}

\usepackage{graphicx}

\newcommand{\R}{\ensuremath{{\mathrm{I\!R}}}}

\newcommand{\dom}{{\rm dom{\,}}}
\newcommand{\hil}{\mathcal H}

\begin{document}

\title{{Dimensionalities of Weak Solutions in  Hydrogenic Systems}}

\author{Alejandro L\'opez-Castillo\\ \small\it Departamento de
Qu\'{\i}mica, Centro Universit\'ario FIEO (UNIFIEO) \\ \small \it Osasco,
SP, 06020-190 Brazil
\\and\\ C\'esar R. de Oliveira\\
\small \it Departamento de Matem\'{a}tica, UFSCar,
S\~{a}o Carlos, SP, 13560-970 Brazil\\ \\}
\date{ \today}
\maketitle

\begin{abstract} A close inspection on the 3D hydrogen atom Hamiltonian
revealed formal eigenvectors  often
discarded in the literature. Although not in its domain,
such eigenvectors belong to the Hilbert space,  and so their time
evolution is well
defined. They are then related to
the 1D and 2D hydrogen atoms and it is numerically found that they have
continuous components, so  that ionization can take place.

\end{abstract}


\vspace{0.5cm}

\noindent Short title: Weak Solutions of the Hydrogen Atom

\clearpage

\section{Introduction}
In order to clearly state the question addressed here, it
is important to recall some
points of the mathematical foundation of observables in quantum mechanics.
There will be two main
contributions, one related to some weak solutions of the Schr\"odinger
equation and other to dimensional
interpretations.

The problem of finding the
correct self-adjoint extension describing the quantum (Schr\"odinger)
operator corresponding to a physical
model can be subtle and difficult. Usually the physicist has a clear
expression for the operator, an
unbounded one acting in a Hilbert space~$\mathcal H$, but it is not
obvious which
domain should be taken (some general references for what follows are
\cite{BEH,RS,Th}).

Let $\langle\psi,\phi\rangle$ denote the inner product in~$\mathcal
H$; if $T$
is a linear operator acting on its dense domain $\dom T\subset \mathcal
H$, then to represent a physical
observable it is necessary that $T$ is hermitian, i.e.,
\[
\langle T\psi,\phi\rangle = \langle \psi,T\phi\rangle, \quad \forall
\psi,\phi\in \dom T.
\]
However this condition is not enough to guarantee that $T$ has real
spectrum and the time evolution it generates is
unitary; the right condition is self-adjointness. The domain of its
adjoin $T^*$ is
\[
\dom T^* = \{\xi\in\mathcal H: \exists \eta\in{\mathcal H} \;{\rm
with}\; \langle \eta,\phi\rangle =
\langle \xi,T\phi\rangle, \; \forall\phi\in\dom T \},
\]
and for $\xi\in\dom T^*,$ one has $T^*\xi=\eta.$ It follows that $T^*$
is well defined if $\dom T$ is dense
in $\mathcal H$, and $T$ is hermitian if, and only if, $T^*$ is an
extension of $T$. The operator $T$ is
self-adjoint if $T=T^*$. Notice also that (often) for bounded operators the
distinction between hermitian and
self-adjoint operators does not exist.

As already mentioned, usually $T$ is hermitian with dense domain, and
one asks if it is also self-adjoint or
has any self-adjoint extension; such extensions are the candidates for
the operator describing the related physical
observable. A nice situation that often occurs, in particular for the
Hamiltonian of the Hydrogen atom (and
other atomic systems as well), is that $T$ is essentially self-adjoint,
i.e., it has just one self-adjoint
extension and the physical operator is well determined. However, there are
situations where there are
infinitely many self-adjoint extensions and each
one should correspond to a different physical circumstance; the choice
is a physical one, not on mathematical bases. Even worse, some hermitian
operators have no self-adjoint
extensions!

The standard example of such framework is the momentum operator
$P=-i\frac{d}{dx}$ for a particle in a box
$[0,1]$. In this case $\mathcal H = L^2[0,1]$, it is natural to take
$\dom P$ as smooth functions
$\psi\in\mathcal H$ such that $\psi(0)=0=\psi(1)$ (so that the particle
remains confined
to the box); the self-adjoint extensions of this hermitian operator are
$P_\alpha$, where $\alpha$ is a complex number with $|\alpha|=1$, and
all elements of $\dom P_\alpha$ satisfy $\psi(1)=\alpha\psi(0)$.

It is worth remarking that if $T$ is hermitian and $\dom T=\mathcal H$,
then $T$ is bounded,
so that in  general such domain questions
are not avoidable. These interesting problems are well explored in the
literature, and as additional
references see
\cite{AGHKH} and for applications to the one-dimensional hydrogen
atom see~\cite{FLM,Gesz}.

Nevertheless, there are some delicate issues in the mathematical
foundations of quantum mechanics that seem
not yet exploited from the physical point of view. The main goal of
this work is to discuss one of
such issues and relate it to a physical situation.

Recall that a self-adjoint Hamiltonian operator
$H$ generates a time evolution $\psi(t)=U(t,0)\psi=e^{-itH}\psi$,
which is a solution of the Schr\"odinger equation
\[
i\frac{d}{dt}\psi(t) = H\psi(t),\quad \psi=\psi(0)\in\dom {H}.
\]
Since $U(t,0)$ is a family of unitary operators, for any time $t$ its
domain is
the
whole Hilbert space $\mathcal H$, so
that it is meaningful to consider $U(t,0)\varphi$ for $\varphi\in\mathcal
H$ but with $\varphi\notin \dom H$,
i.e., the time evolution is not
restricted to the domain of $H$. Sometimes $U(t,0)\varphi$, for $\varphi\notin \dom H$, is called a
weak solution of the Schr\"odinger
equation.

An unusual situation will be presented. A self-adjoint  operator $H$ with
dense
$\dom H\subset\hil=L^2(\R^3)$ will be considered,
 vectors
$\Xi\in\hil$ not belonging to its domain will be given, although they are
pseudo-eigenvectors of $H$, that is,
\begin{equation}\label{pseudoeigenvalue1}
H\Xi = \lambda_\Xi \Xi,
\end{equation}
for $\lambda_\Xi\in \R$. Some numerical calculations will indicate
that $\Xi$ has a nonzero component in
the continuous subspace of $H$, so that the na\"{\i}ve time evolution
built from (\ref{pseudoeigenvalue1}) gives
an incorrect answer. It will be argued that such solutions are related to
the same model in smaller
dimensions. Furthermore, the physical system in question is one of the
most celebrated models in
quantum mechanics, the three dimensional (3D) hydrogen atom.

\section{Pseudo-Eigenvectors as Weak Solutions}

The hermitian Hamiltonian of the 3D hydrogen atom is
\[
H_0=-\frac{\hbar^2}{2\mu}\Delta - \frac{e^2}{r},\quad \dom
H_0=C^\infty_0(\R^3)\subset L^2(\R^3),
\]
where $\mu$ is the electron mass, $e$ its electric charge and  $C^\infty_0(\R^3)$ denotes the set of
smooth functions with compact
support. This operator is essentially
self-adjoint and its unique self-adjoint extension
$H_H$, {\it the} 3D hydrogen atom
Hamiltonian, reads
\begin{equation}\label{HH}
H_H=-\frac{\hbar^2}{2\mu}\Delta - \frac{e^2}{r},\quad \dom
H_H=H^2(\R^3),
\end{equation}
with $H^2(\R^3)$ denoting an appropriate Sobolev space; in particular,
$H^2(\R^3)$ is a subspace of $L^2(\R^3)$,
it is also the natural domain of the free particle Hamiltonian and all
its
elements are continuous functions~\cite{RS}.

The usual spectral analysis of $H_H$ can be performed and its well-known
eigenvalues
\[
-\frac{\mu e^4}{2 \hbar^2 n^2},\quad n\ge1,
\]
can be found. Recall that the closed subspace $\hil_p$ generated by its
eigenvectors is named the point
subspace of
$H_H$ and its orthogonal complement $\hil_{ac}$ is a nontrivial subspace (i.e., it has nonzero elements) and
named the absolutely continuous (or scattering) subspace of
$H_H$. Physically, the members of $\hil_p$ are the bound states while
the
elements of $\hil_{ac}$ describe the
ionizing atomic states (this interpretation follows, for instance, by the
RAGE Theorem
\cite{CFKS,deOdoC}).

The eigenvalue equation for the 3D hydrogen atom Hamiltonian is separable
in spherical
coordinates $r\ge0,\, 0\le \theta\le\pi,\,0\le \phi\le2\pi$, and by taking the standard representation
\begin{equation}\label{RTP}
\Psi(r,\theta,\phi)=R(r)\Theta(\theta)\Phi(\phi),
\end{equation}
 the equation for
$\Theta(\theta)$ is given by~\cite{pauling}
\begin{equation}\label{ThetaEq}
\frac{1}{\sin\theta}\frac{d}{d\theta}\left(\sin\theta
\frac{d\Theta}{d\theta}\right) -\left(\frac{m^2}{\sin^2\theta}-\ell(\ell+1)
\right)\Theta=0,
\end{equation}
with $m$ and $\ell\ge0$ being integer constants. For each $\ell$ value one
has
$-\ell\le m\le \ell$.

Consider first  the particular
case  $\ell=0$; it follows that $m=0$ and (\ref{ThetaEq}) reduces to
\begin{equation}\label{danadaEq}
\frac{1}{\sin\theta}\frac{d}{d\theta}\left(\sin\theta
\frac{d\Theta}{d\theta}\right) =0.
\end{equation}
The usual normalized solution of this equation is
$\Theta_{0,0}(\theta)=\Theta_{l=0,m=0}({\theta})=1/\sqrt2$. However, there
is also
the additional solution (that will play a major role here)
\begin{equation}\label{qsi}
\xi_{0,0}(\theta) = \frac{\sqrt6}{\pi}\ln\left[ \tan\left(
\frac{\theta}{2}\right) \right].
\end{equation}
This is just one instance of
additional solutions $\xi_{\ell,m}$ of
(\ref{ThetaEq}) for $\ell,m$ as above; such
solutions are Legendre function of the second kind~\cite{arfken,math}.
The $\xi_{\ell,m}$ solutions have been discarded in the mathematical
literature
since
they are not continuous at
$\theta=0$ and $\theta=\pi$, and so via (\ref{RTP}) they do not generate
elements in the domain of
$H_H$; and discarded in the physical literature~\cite{pauling} by
arguing they are not bounded functions.

By taking the usual radial $R_{n,\ell}(r)$ and azimuthal $\Phi_m(\phi)=\frac
1{\sqrt{2\pi}}e^{im\phi}$ solutions for the 3D hydrogen
atom, set ($\ell<n$)
\[
F_{n,\ell,0}(r,\theta,\phi)= \left[\left(\frac{2}{n a_0}\right)^3
\frac{(n-\ell-1)!}{4 \pi n \left[(n+\ell)!\right]^3} \right]^\frac12
\exp\left(-\frac{r}{n a_0}\right)\left(\frac{2 r}{n a_0}\right)^\ell
L^{2 \ell+ 1}_{n+\ell}\left(\frac{2 r}{n a_0}
\right)
\] so that one gets the
standard eigenfunctions (here restricted to $m=0$)
\[
\Psi_{n,\ell,0}(r,\theta,\phi):= F_{n,\ell,0}(r,\theta,\phi)
\Theta_{\ell,0}(\theta),
\]
and now the additional ones

\[
\Xi_{n,\ell,0}(r,\theta,\phi):=F_{n,\ell,0}(r,\theta,\phi)
\xi_{\ell,0}(\theta),
\]
where $a_0=\hbar^2/(\mu e^2)$ denotes the Bohr radius and
$L^{2\ell+1}_{n+\ell}$ are the
Laguerre polynomials.

For $m\ne0$ the probability density generated by $\xi_{\ell,m}$
 diverges (recall the Jacobian is
$r^2 \sin\theta$), i.e.,
\[
\int_0^\pi \sin\theta\, |\xi_{\ell,m\ne0}(\theta)|^2\, d\theta =\infty;
\]
thus the corresponding functions $\Xi_{n,\ell,m\ne0}(r,\theta,\phi)$  do not
belong to the Hilbert space
$L^2(\R^3)$. Hence, it is meaningless to talk about their
time evolution even as weak
solutions of the 3D hydrogen atom Schr\"odinger equation.

However, for $m=0$ the probability density generated by $\xi_{\ell,0}$
does not diverge, i.e., by choosing
appropriate constants it is found that
\[
\int_0^\pi \sin\theta\, |\xi_{\ell,0}(\theta)|^2\, d\theta =1,
\]
so that together with the  $R(r)$ and $\Phi(\phi)$ counterparts in
(\ref{RTP}),
$\xi_{\ell,0}$ generates elements $\Xi_{n,\ell,0}$  of
$L^2(\R^3)$, as given by the expression above.

Notice that although $\Xi_{n,\ell,0}$ does not belong to the domain of
$H_H$,
one formally finds
\begin{equation}\label{pseudoeigenvalue}
H_H\Xi_{n,\ell,0} = \lambda_{n} \Xi_{n,\ell,0},\quad \lambda_{n} =
-\frac{\hbar^2}{2\mu a_0^2 n^2},\quad \forall n, 0\le \ell<n,
\end{equation}
so that $\Xi_{n,\ell,0}$ and $\lambda_{n}$ are pseudo-eigenvectors and
pseudo-eigenvalues of $H_H$,  respectively. Another
point supporting the use of the adjective ``pseudo'' is that
$\Xi_{n,\ell,0}$
are not orthogonal to
every $\Psi_{n,\ell',0}$ (for instance, $\Xi_{n,0,0}$ is not orthogonal to
 $\Psi_{n,\ell',0}$ with odd~$\ell'$).

The question to be addressed now is about the time evolution of
$\Xi_{n,\ell,0}$, which is well defined since $\Xi_{n,\ell,0}\in L^2(\R^3)$
and so $U(t,0)\Xi_{n,\ell,0}$ is a weak solution of the 3D hydrogen atom
Schr\"odinger equation
\[
i\hbar\frac{\partial}{\partial t}\psi = H_H \psi.
\]
Based on (\ref{pseudoeigenvalue}) the na\"{\i}ve expression for the solution $\Xi_{n,\ell,0}(t)=e^{-i H_H
t/\hbar}\Xi_{n,\ell,0}$ is
\begin{equation}\label{naivesolution}
\Xi_{n,\ell,0}(t)=e^{-i \lambda_{n} t/\hbar}\Xi_{n,\ell,0};
\end{equation}
such solution is not correct for $\Xi_{n,\ell,0}$ is not in the
domain of $H_H$; if (\ref{naivesolution})
holds then $\Xi_{n,\ell,0}(t)$ would be strongly differentiable and so
one could conclude that
$\Xi_{n,\ell,0}\in\dom H_H$. Notice that (\ref{naivesolution}) is correct
if $\Xi$ is replaced by
$\Psi$.

From the dynamic point of view it is important to check if
$\Xi_{n,\ell,0}$ belongs to the point subspace $\mathcal H_p$
associated to $H_H$ or if it has a component in the absolutely continuous
subspace $\mathcal H_{ac}$. In the latter case ionization can take place; in the former case these
generalised eigenstates are written as superpositons of ordinary eigenstates, even if they do not belong to
the domain of the operator, and so such solutions would be, in some sense, between linear  combination of
ordinary eigenvectors and the continuous space (but with no ionization).

\begin{figure}
\begin{center}
\vskip 10pt
\includegraphics[width=6cm, height=6cm]{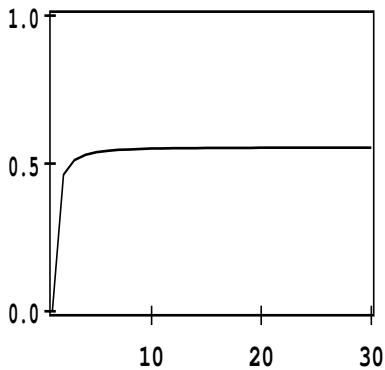}
\caption{$P$ as function of $N$.}
\label{FigPN}
\end{center}
\end{figure}

In order to check if $\Xi_{n,\ell,0}$
is generated by $\Psi_{n',\ell',m}$, $n'\ge1$,
$\ell'=0,\cdots,n'-1,$
$-\ell'\le m\le \ell'$, i.e., if $\Xi_{n,\ell,0}$ belongs to the point
subspace of
$H_H$, consider
\[
\Xi_{n,\ell,0} = \sum_{n'=1}^\infty \sum_{l'=0}^{n'-1} C_{n',l'}^{(n)}\,
\Psi_{n',l',0} + \chi^c_{n,\ell},
\]
with $\chi^c_{n,\ell}$ denoting the component of $\Xi_{n,\ell,0}$ in the
continuous subspace
$\mathcal H_{ac}$. Notice that clearly
$\Xi_{n,\ell,0}\perp \Psi_{n',\ell',m}$ if $m\ne0$. It was numerically found
that
$\|\chi^c_{1,0}\|>0.8$, as indicated in figure~\ref{FigPN} for $n=1$ (in  figure~\ref{FigPN} the
values of $P(N)$  are exact, since symbolic calculus was used); the parameter $P(N)^2\equiv\sum_{n'=1}^N
\sum_{l=0}^{n'-1} |C_{n',l}^{(1)}|^2$ is an approximation for
$1-\|\chi^c_{1,0}\|^2$. Similar results were found for other values
of~$\ell$.

Therefore one concludes that $\Xi_{n,\ell,0}$ have both nonzero point and
continuous components, so that their time
evolutions actually are not described by (\ref{naivesolution}), but give
 nonzero
probabilities $\|\chi^c_{n,\ell}\|^2$ of ionization (and then far from being
bound states).

\section{Lower Dimensional Hydrogen Atom}
\begin{figure}
\begin{center}
\vskip 10pt
\includegraphics[width=6cm, height=6cm]{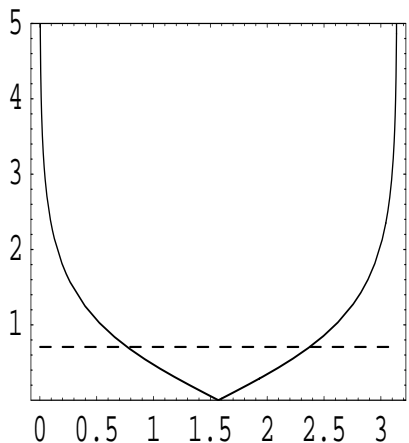}
\caption{$|\xi_{0,0}(\theta)|$ and  $|\Theta_{0,0}(\theta)|$ (dashed) as
function of~$\theta$.}
\label{Figzerozero}
\end{center}
\end{figure}

\begin{figure}
\begin{center}
\includegraphics[width=12cm, height=8cm]{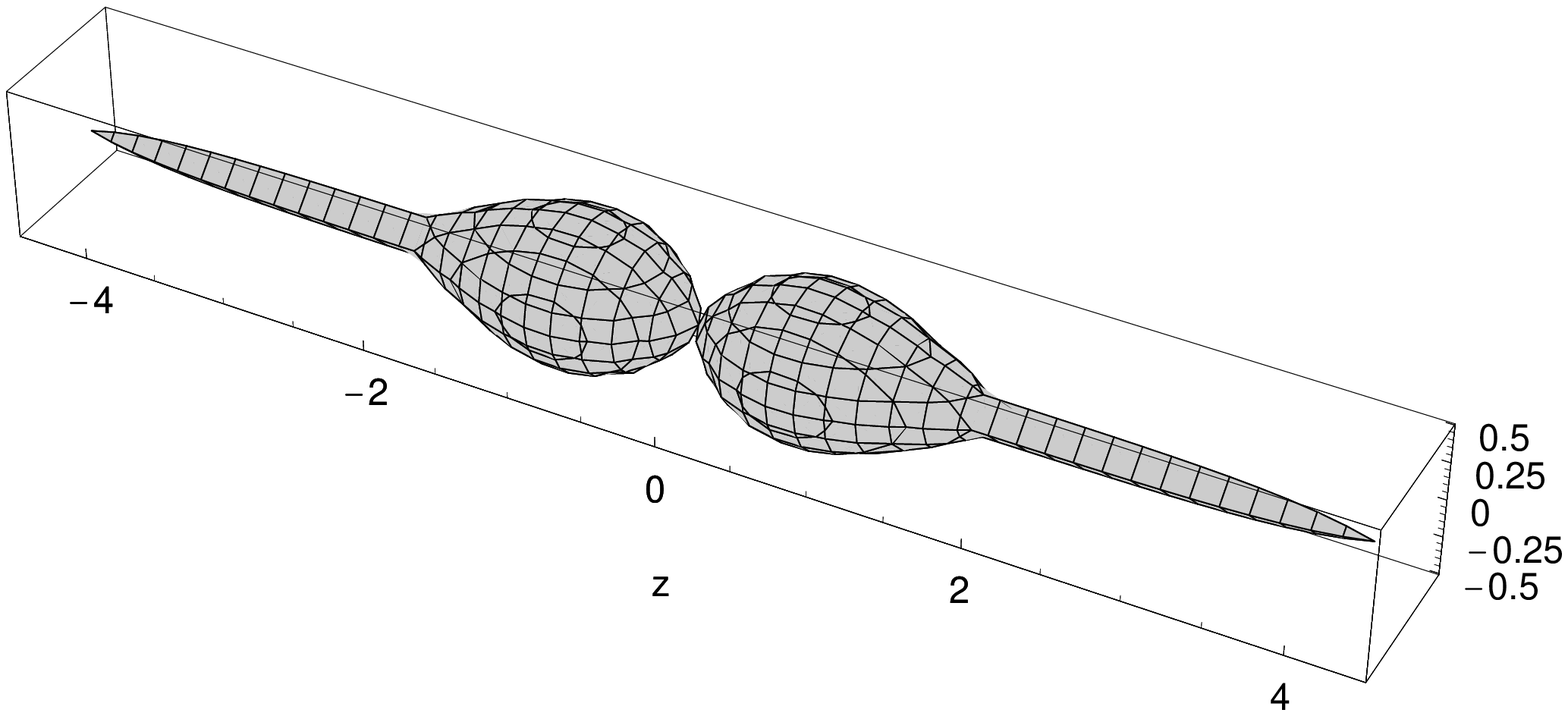}
\caption{A boundary surface of $|\Xi_{n=1,\ell=0,0}(r,\theta,\phi)|^2$. }
\label{function3D}
\end{center}
\end{figure}

The fact that all $\Xi_{n,\ell,0}$ belong the Hilbert space raises the possibility of finding physical
meanings  for them; this section aims at discussing possible physical contents of these pseudo-eigenvectors.
The first crucial remark is that formally $\Xi_{n,\ell,0}$ has null azimuthal angular
momentum ($m=0$). The  solution $\Xi_{n,0,0}$ has also null total angular
momentum (both
$\ell=0=m$), but there is a lack of rotational symmetry (see~(\ref{qsi}));
this particular solution gives
a clue on the physical interpretation. In fact, in comparision with ordinary eigenfunctions
$\Xi_{n,\ell,0}(r,\theta)$ 
are elongated over the $z$-axis with a logarithmic divergence at
$\theta=0$ and $\pi$.
Figure~\ref{Figzerozero} shows the absolutely values of
$\xi_{0,0}(\theta)$ and $\Theta_{0,0}(\theta)$ as a
function of~$\theta$, and Figure~\ref{function3D} a boundary surface of the 3D wavefunction $\Xi_{1,0,0}$,
which is to be compared with $\Psi_{0,0,0}$ that has complete radial symmetry (its
boundary surfaces are spheres centrered at the origin). Hence, there is a strong indication that
$\Xi_{n,0,0}$ are reminiscent of classical
trajectories performing one-dimensional (1D) like motion, in agreement with
its null angular momentum and lack
 of rotational symmetry.  So, it is natural to relate such
wavefunctions to the 1D hydrogen atom, an interesting and
controversial subject, popularized by the
work of Loudon \cite{Lou} published in 1959.

Loudon stated that the 1D hydrogen atom was twofold degenerate, having
even and odd eigenfunctions for each
eigenvalue, except for the (even) ground state having infinite binding
energy. Typically 1D systems have no
degenerate eigenvalues, and Loudon justified the double degeneracy as a
consequence of the singular atomic
potential. Andrews \cite{And1} questioned the existence of a ground
state with infinite binding energy. Ten
years later Haines and Roberts \cite{HR} revised Loudon's work and
obtained that the even wave functions,
with continuous eigenvalues, were complementary to odd functions, but
such results were criticized by
Andrews \cite{And2}, who did not accepted the continuous eigenvalues.
Gomes and Zimerman \cite{GZ} argued
that the even states with finite energy should be excluded. Spector and
Lee \cite{SL} presented a
relativistic treatment that removed the problem of infinite binding
energy of the ground state. Several other
works \cite{DPST,BKB,NVS,LC,OL,FLM,XDD,LL} (see also references therein)
have discussed this (apparent)
simple problem.

The 1D hydrogen atom has been used as a simplification of the 3D model
in several theoretical and
numerical studies \cite{JSS,DKS,LCO3}. It is then interesting that Cole
and Cohen
\cite{CC} and Wong et al.\ \cite{W} have
reported some experimental evidence for the 1D hydrogen atom. The
``quasi-1D'' solutions $\Xi_{n,0,0}$ are
natural candidates to describe such experimental observations and may
be relevant for an appropriate
justification for the use of 1D simplifications. Lastly, the
1D eigenvalues coincide with the
eigenvalues of the 3D hydrogen model.

\begin{figure}
\begin{center}
\vskip 10pt
\includegraphics[width=6cm, height=6cm]{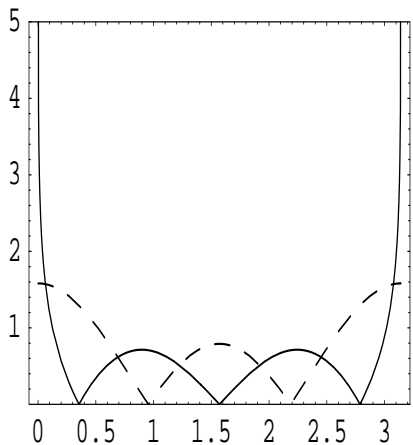}
\caption{$|\xi_{2,0}(\theta)|$ and $|\Theta_{2,0}(\theta)|$ (dashed) as
function of $\theta$.}
\label{FigDoisDois}
\end{center}
\end{figure}

Now the solutions $\Xi_{n,\ell\ne0,0}$ have nonzero  total  angular
momentum while zero angular
momentum in  the $z$-direction, and the logarithm divergence for
$\theta=0$ and~$\pi$ is also present for all
$\xi_{l,0}$, indicating that the $z$-axis plays a special role in the
classical trajectories analogy. So it
is possible to interpret that $\Xi_{n,\ell\ne0,0}$ is related to
two-dimensional motions taking place in planes
containing the $z$-axis, i.e., to the 2D hydrogen atom.
Figure~\ref{FigDoisDois} illustrates such
interpretation for
$\ell=2,m=0$. The
2D hydrogen atom has also been considered in the literature (see
\cite{YGCWC,delCV,RR,VP,PP}
and references therein), but its history is not as controversial as for
the 1D case.

Finally, a word about $\Xi_{n,\ell,m\ne0}$; since they do not belong to
the Hilbert space, based on the
above discussion and proceeding heuristically, it is tempting to interpret such solutions as the
``contribution'' due to the classical
trajectories which come into collision with the nucleus, and the
mathematical apparatus prudently avoids them
explicitly (maybe a mathematical consequence of the uncertainty principle).

\section{Conclusions}
One is naturally inclined to presume that higher dimensional quantum
models carry somehow lower
dimensional dynamics, and the study of such simpler models could mimics
important aspects of
the original one. Of course, in general the difficulties of performing
such dimensional reductions are enormous, and usually carried out by
``brute force.''

The case of the 3D hydrogen atom discussed in this
work has revealed a particular and
interesting framework: there are experimental evidence for the 1D
hydrogen atom; the 3D hermitian model has just one self-adjoint extension,
and its
 3D eigenvalue equation
presents formal solutions $\Xi_{n,\ell,0}$ that do not belong to the
domain of the corresponding Hamiltonian
operator; in spite of being formal eigenvectors, these solutions live in
the underlying Hilbert space
and present a component in the continuous subspace
of the Hamiltonian so that, for an electron in such state, ionization can
take place;
  these solutions have formally
zero azimuthal angular momentum, with integrable probability
densities, and are concentrated around the $z$-axis,  indicating their 1D
and 2D
character for $\Xi_{n,0,0}$ and $\Xi_{n,\ell\ne0,0}$, respectively.
Summing up, such solutions are
reminiscent of 1D and 2D classical trajectories and give a connection
between the hydrogen atom in different
dimensions.

How general is this framework? This is a fascinating open question, whose
answer could
eventually improve the interpretations.

In addition, notice that there is an attractive relation between the
dimensional interpretations advocated in
this work and the mathematical formalism, which exhausts the possibilities for (pseudo-)eigenvectors. For
genuine 3D motion it presents eigenfunctions in the Hamiltonian
operator domain; for 1D and 2D reminiscent trajectories it presents
eigenfunctions in the Hilbert space but not
in the domain of the operator; and for those colliding trajectories (axial
divergence)  the
formal eigenfunctions do not belong
to the Hilbert space.

\subsubsection*{Acknowledgments} {\small AL-C thanks FAPESP. CRdeO thanks
the partial
support by CNPq}.

\end{document}